\documentstyle[twocolumn,aps,epsf]{revtex}
\def\theequation{\arabic{section}.\arabic{equation}}
\def\bF{{\bf F}}
\def\bj{{\bf j}}

\def\br{{\bf r}}
\def\bu{{\bf u}}
\def\bv{{\bf v}}

\def\bpart{\bar\partial}
\newcommand{\be}{\begin{equation}}
\newcommand{\ee}{\end{equation}}
\begin{document}
\draft
\title{Vortex dynamics in the two-fluid model}
\author{D. J. Thouless,$^1$ M. R. Geller,$^{2}$ W. F. Vinen$^3$,
J.-Y. Fortin$^4$, and S. W. Rhee$^1$}
\address{$^1$Department of Physics, Box 351560, University of Washington,
Seattle, Washington 98195}
\address{$^2$Department of Physics and Astronomy, University of Georgia,
Athens, Georgia 30602}
\address{$^3$School of Physics and Astronomy, University of Birmingham,
Birmingham B15 2TT, England}
\address{$^4$CNRS Laboratoire de Physique Th\'eorique,
 Universit\'e Louis Pasteur, 67084 Strasbourg Cedex, France}
\date{\today}
\maketitle

\begin{abstract}

We have used two-fluid dynamics to study the discrepancy between the
work of Thouless, Ao and Niu (TAN) and that of Iordanskii.  In TAN no
transverse force on a vortex due to normal fluid flow was found,
whereas the earlier work found a transverse force proportional to
normal fluid velocity $u_n$ and normal fluid density
$\rho_n$.  We have linearized the time-independent two-fluid equations
about the exact solution for a vortex, and find three solutions which
are important in the region far from the vortex.  Uniform superfluid
flow gives rise to the usual superfluid Magnus force.  Uniform normal
fluid flow gives rise to no forces in the linear region, but does
not satisfy reasonable boundary conditions at short distances.  A
logarithmically increasing normal fluid flow gives a viscous force.
As in classical hydrodynamics, and as in the early work of Hall and
Vinen, this logarithmic increase must be cut off by nonlinear effects
at large distances; this gives a viscous force proportional to
$u_n/\ln u_n$, and a transverse contribution which goes like $u_n/(\ln
u_n)^2$, even in the absence of an explicit Iordanskii force. In the
limit $u_n\to 0$ the TAN result is obtained, but at nonzero $u_n$
there are important corrections that were not found in TAN.  We argue
that the Magnus force in a superfluid at nonzero temperature is an
example of a topological relation for which finite-size corrections
may be large.
\end{abstract}

\setcounter{equation}{0}
\section{Introduction}\label{sec:int}

In recent years papers have been published which appear to give an exact
expression for the transverse force, including the superfluid Magnus
force, acting on a quantized vortex moving in a neutral superfluid
\cite{tan96,wexler97}.  These methods were extended to give the transverse
force on a flux line in a superconductor in the ultraclean limit
\cite{geller98}, and to give a corresponding (but not easily calculable)
expression for the longitudinal force on a vortex in a neutral superfluid
\cite{tang98}.  These results have been criticized
\cite{volovik96,hallhook98,sonin98} on the grounds that they do not agree
with the widely, but by no means universally, accepted results in the
literature \cite{sonin97,donnelly91}.  In particular, a transverse force
proportional to the superfluid density, and independent of the normal
fluid density and velocity, was found, whereas it is generally supposed
that the asymmetrical scattering of phonons or rotons by a vortex,
demonstrated many years ago by Lifshitz and Pitaevskii \cite{pitaevskii58}
and by
Iordanskii \cite{iordanskii64}, should lead to a contribution to the
transverse force proportional to the normal fluid density and to the
velocity of the vortex relative to the normal fluid component.

In order to shed new light on this controversy we have decided to 
examine carefully the predictions of the two-fluid model for a pinned 
vortex in a moving fluid.  The two-fluid equations should be valid 
under conditions in which the pressure, temperature, and normal and 
superfluid velocities are varying slowly on the length scale set by 
the normal fluid mean free path.  This cannot be true close to the 
core of a vortex, but it can be true everywhere if the vortex is 
pinned by a sufficiently large cylinder, on which either zero velocity 
or zero rate of shear (slip) boundary conditions are imposed.  The 
equations are also valid at sufficient distances from the pinning 
center, even if the pinning center has a radius less than the mean
free path.  This line of thought is not entirely new, since similar
arguments were used in the original work of Hall and Vinen
\cite{hallvinen56} on vortex arrays in helium, and by Vinen
\cite{vinen57} on mutual friction.  Their arguments, however, did not
address the question of a transverse force.

In sec.\ \ref{sec:equilib} we write down the equations of two-fluid 
theory for steady state flow, and examine the simple equilibrium 
solution associated with a vortex in a stationary fluid.

In sec.\ \ref{sec:linear} we set up linearized steady state equations 
and study their solutions.  We want to find solutions linearized about 
the equilibrium vortex solution, and examine the behavior of the 
steady state solutions at large distances from the vortex.  Such an 
approach should be adequate to derive any transverse force 
proportional to normal fluid velocity and superfluid circulation, such 
as is quoted in refs.\ \ref{ref5}, \ref{ref6} and \ref{ref7}, as well 
as to derive the force proportional to superfluid velocity and 
superfluid circulation.  The partial differential equations are
separable in cylindrical polar coordinates, so by a combination of
analytical and numerical methods we can get a fairly complete
description of those solutions that have the same symmetry as uniform
flow past the vortex.

In sec.\ \ref{sec:bc} we examine the influence of boundary conditions on
the solutions of the equations of motion.  In our initial discussion we
look at the case of a cylindrical inner boundary on which the normal
fluid is at rest and the temperature is uniform.  There is a
complication, known long ago for the classical theory of  flow of
viscous fluid past a cylindrical obstacle, and discussed for superfluids
in the works of Hall and Vinen \cite{hallvinen56} and Vinen
\cite{vinen57}, which is that the  linearized equations are not valid at
very large distances, where the  Reynolds number is of order unity even
though the speed of flow is small  \cite{batchelor67}.  This is true
for any reasonable set of short-distance boundary conditions.  This
difficulty was circumvented by Oseen, and we can use the same
technique.  The drag force, in the linear regime, is transmitted by a normal fluid 
velocity which increases with the logarithm of the distance from the 
vortex core, but this must be cut off at some distance $R_c$ at which 
nonlinear inertial effects become more important than the viscous 
force. The cut-off may alternatively be set by the spacing between 
vortices, by the finite size of the system, or by finite viscous 
diffusion length at nonzero frequency of vortex motion. It was 
initially surprising to us that for these conventional boundary 
conditions there should be a transverse force, but it exists because 
the coupling between normal fluid flow and superfluid flow at small 
distances from the inner boundary produces it.  We find that the 
magnitude of this transverse force is proportional to 
\be\label{eq:tnforce} 
\rho_n\frac{h}{m}\frac{u_n}{[\ln(R_c/r_0)]^2}\frac{L^2}{r_0^2} \;, \ee
where $u_n$ is the normal fluid velocity at distances of the order 
$R_c$, $r_0>>L$ is the distance from the axis at which inner boundary 
conditions are applied, and $L$ is the mean free path for excitations.  
This transverse force due to the normal fluid is small unless the 
radius of the boundary is comparable with the mean free path.  A 
significant result that we get from our generalization of Lamb's 
solution \cite{lamb45} of the Oseen equation is that at large distances 
from the vortex there is a normal fluid circulation, contrary to what 
was assumed by TAN, but this circulation is proportional to $1/\ln^2U$, 
where $U$ is the asymptotic normal fluid velocity, so it goes slowly to 
zero as the normal fluid velocity goes to zero.  Our results do agree 
with TAN's relation between the transverse force and the circulation of 
momentum at large distances. Some details of the calculation are shown 
in Appendix \ref{sec:firstorder}.

In sec.\ \ref{sec:iordanskii} we give a brief discussion of how the 
inner boundary conditions can be modified to allow for the transverse 
force generated by flow of normal fluid past a bare vortex.  Our study 
of the asymptotic flow shows that any force dependent on the normal 
fluid velocity is modified by a logarithmic dependence on the cut-off
distance $R_c$.  In the  low velocity limit any force due to the normal 
fluid motion, such as the Iordanskii \cite{iordanskii64} force, 
is renormalized away by logarithmic denominators; the force on a bare 
vortex has the form given by eq.\ (\ref{eq:tnforce}) in the case 
$r_0\approx L$. In  practice these terms should not be ignored, 
and may give a significant normal fluid contribution to the 
transverse force on a 
vortex, in addition to the superfluid Magnus force.  The relation 
between vortex velocity and the transverse force on it does have a 
topological structure similar to the relation between voltage and 
current in a quantum Hall device \cite{thouless99}, but, unlike the 
quantum Hall conductance, this relation is subject to large corrections 
due to the finite size of the system, or the nonzero value of the 
velocity, decreasing only as the reciprocal of the square of the 
logarithm of the relevant length parameter.

There is a discussion of the relation of this work to the original
work of Hall and Vinen in Appendix \ref{sec:hallvinen}.

In a recent paper by Sonin \cite{sonin00} there is a similar discussion
of the relation of the TAN results to two-fluid theory.  Sonin argues that
there must be an additional contribution to the Berry phase from the flow of
phonons or rotons past a vortex, and that this force is transmitted to large
distances by the usual hydrodynamic processes.

\section{Equilibrium solutions}\label{sec:equilib}
\setcounter{equation}{0}

The equations of two-fluid hydrodynamics are given in the book by
Khalatnikov \cite{khalat65}, in eqs.\ 9-13, 9-14, 9-15, 9-16, 8-14, 
8-19$'$ and 9-11. We want these equations in steady state conditions, 
when all time derivatives are zero.  The equation of conservation of 
matter is then
\be\label{eq:matter}
\nabla\cdot(\rho_s \bv +\rho_n \bu) =0\;, 
\ee
where we have written $\bv$ for the superfluid velocity, and $\bu$ for 
the normal fluid velocity.  The momentum balance equation 
(Navier--Stokes equation) is
\[
\nabla_k\bigl(\rho_s v_iv_k+\rho_n u_iu_k\bigr)+\nabla_i p =\nabla_k 
[\eta (\nabla_k u_i +\nabla_i u_k)] \]
\be\label{eq:forces}
-\nabla_i\Bigl[\bigl({2\over 3}\eta+ \zeta_1\rho
-\zeta_2\bigr) \nabla\cdot \bu+\zeta_1\bu\cdot \nabla \rho\Bigr]
\;, \ee
where $\eta$ is the first viscosity, and the $\zeta$s are coefficients 
of second viscosity in the two-fluid model. We have set $\nabla\cdot 
\bj$ equal to zero here, from eq.\ (\ref{eq:matter}).
The equation for superfluid dynamics integrates to give
\be\label{eq:super}
\mu-\mu_0+{1\over 2} v^2=(\zeta_4-\rho\zeta_3)\nabla\cdot \bu- \zeta_3
\bu\cdot\nabla\rho\;.
\ee
The equation for thermal balance is
\be\label{eq:heat}
\nabla\cdot\Bigl(S\bu-\kappa{\nabla T\over T}\Bigr)={R\over T}\;,
\ee
where $S$ is the entropy density, $\kappa$ is the thermal 
conductivity, and $R$ is the rate of heat generation by dissipative 
processes, a term which is quadratic in the normal fluid velocity and 
temperature gradient, and will be set equal to zero in the linear 
approximation.  Finally, there is a thermodynamic relation
\be\label{eq:thermo}
dp= SdT+\rho d\mu+{1\over 2}\rho_nd(\bv-\bu)^2\;.
\ee

The equilibrium vortex, if the pinning center has cylindrical 
symmetry, is described very simply by the solution
\[
\bu=0\;,\ \ \bv={\hbar\over mr}\hat \theta\;,\ \ \mu=\mu_0-
{\hbar^2\over 2m^2r^2}\;,\]
\be\label{eq:equil} {dp\over dr}=\rho_s(r){\hbar^2\over 
m^2r^3}\;,\ \ T=T_0\;.
\ee
This is compatible with the local thermodynamic relation, eq.\
(\ref{eq:thermo}).

\section{Linearized steady state equations}\label{sec:linear}
\setcounter{equation}{0}

Now we want to linearize the equations 
(\ref{eq:matter})--(\ref{eq:thermo}) about the vortex solution 
(\ref{eq:equil}).  We will take full account of first order terms in 
$\delta \bv, \bu, \delta p, \delta T, \delta \mu$, but will neglect 
the consequent variations of $\rho_s, \rho_n, S$, and of the
transport coefficients.  This is an additional simplifying assumption, 
and does not affect the asymptotic form of the solutions, since such 
variations all come into the equations accompanied by an additional 
factor of $r^{-2}$. We consider only motion in the $xy$ plane, so all 
equations will be two-dimensional.  The equation of conservation of 
matter becomes
\be\label{eq:l2matter}
\rho_n\nabla\cdot \bu=-\rho_s \nabla^2w  \;,
\ee
where the superfluid velocity has been written as
\be\label{eq:spotential} \bv=\nabla (W+w)\;,\ \ {\rm where}\ \ 
W=\hbar\theta/m \;. \ee
It is convenient now to write the normal fluid velocity in the form
\be\label{eq:curlgrad}
u_i= \epsilon_{ij}\nabla_j\chi -{\rho_s\over\rho_n}\nabla_i w  \;,
\ee
where $\epsilon_{ij}$ is the alternating tensor in two dimensions.
 The linearized Navier--Stokes equation can then be written
\[\nabla_i\Bigl(\rho_s(\nabla_jW)\nabla_jw +\delta p\Bigr) +\rho_s
(\nabla_iW)\nabla^2w\]
\be\label{eq:l2forces}
=\eta\epsilon_{ij}\nabla_j\nabla^2\chi-{\rho_s\over\rho_n} 
\bigl({4\over
3}\eta- \zeta_1\rho +\zeta_2\bigr) \nabla_i(\nabla^2 w)  \;.  \ee
This can conveniently be split into two parts by taking the curl and
divergence of the equation.  The curl gives
\be\label{eq:l2curl} \rho_s\epsilon_{ij}(\nabla_iW)\nabla_j\nabla^2 w=
\eta\nabla^4 \chi \;,\ee 
which is an equation relating the viscous force on the normal fluid to 
the momentum flow in the superfluid.
The divergence gives
\[ \nabla^2 \delta p+2\rho_s
\nabla_i\nabla_j[(\nabla_iW)\nabla_jw]\] 
\be\label{eq:l2div} +{\rho_s\over\rho_n}\bigl({4\over
3}\eta- \zeta_1\rho +\zeta_2\bigr) (\nabla^4 w)=0  \;.  \ee
The superfluid equation gives
\be\label{eq:l2super}
\delta \mu+(\nabla_iW)\nabla_iw
=-{\rho_s\over\rho_n}(\zeta_4-\rho\zeta_3)\nabla^2w\;.
\ee
The heat flow equation gives
\be\label{eq:l2heat}
{S\rho_s\over\rho_n}\nabla^2w +{\kappa\over T}\nabla^2 \delta T=0\;.
\ee
The linearized thermodynamic relation gives
\[ \delta p= S\delta T+\rho \delta\mu+ \rho(\nabla_iW)\nabla_i w \]
\be\label{eq:l2thermo} -\rho_n\epsilon_{ij}(\nabla_iW)\nabla_j \chi\;.
\ee

Substitution of eqs.\ (\ref{eq:l2thermo}),  (\ref{eq:l2heat}), 
(\ref{eq:l2super}), in eq.\  (\ref{eq:l2div}) gives
\[ 2\rho_s \nabla_i\nabla_j[(\nabla_iW)\nabla_jw]
-\rho_n\nabla^2 [  \epsilon_{ij}(\nabla_iW)\nabla_j \chi] \]
\[  +{\rho_s\over\rho_n}\bigl({4\over
3}\eta- \rho\zeta_1-\rho\zeta_4 +\zeta_2+\rho^2\zeta_3\bigr) (\nabla^4
w) \]
\be\label{eq:l2div2} -{S^2T\rho_s\over \kappa\rho_n} \nabla^2w=0  \;. 
\ee 
The last two terms on the left side of this equation describe 
counterflow between normal and superfluid components, while the first 
two terms give coupling of this counterflow to the vortex motion.
From now on we will ignore the contributions of  second viscosity,  which
greatly simplifies the analysis.  Only for $\zeta_2$ do we have experimental
values (from the observed attenuation of first sound) \cite{wilks67},  
although we know from Khalatnikov's \cite{khalat65} eq. (9.12) and the 
inequality (9.18) that their contribution to the coefficient of $\nabla^4 
w$ is positive.  With this simplification we  can rewrite eqs.\ (\ref{eq:l2curl}) and
(\ref{eq:l2div2}) in a  dimensionless form. These coupled linear fourth
order equations have  eight independent solutions for each angular symmetry.

Since temperature gradient and heat flow may be important both in the 
interpretation of these equations and in the boundary conditions, we 
take the gradient of eq.\ (\ref{eq:l2thermo}), and substitute eqs.\ 
(\ref{eq:l2forces}) and (\ref{eq:l2super}) to get
\[
S\nabla_iT= -\rho_s(\nabla_i W)\nabla^2w-\rho_s\nabla_i[(\nabla_jW) 
\nabla_j w] \]\[ -{\rho_s\over\rho_n}{4\over 3}\eta \nabla_i\nabla^2w
+\eta\epsilon_{ij}\nabla_j\nabla^2 \chi \]\be\label{eq:l2temp}
 + \rho_n\nabla_i 
[\epsilon_{jk}(\nabla_jW)\nabla_k\chi] \;.\ee

We introduce the length scale $L$ defined by
\be\label{eq:length}
L^2={4\eta\kappa\over 3S^2T}\;.
\ee
From kinetic theory $L$ should be of the order of the mean free path 
for the excitations.  We rewrite the equations in terms of the 
dimensionless complex variables $z=(x+iy)/L$ and its conjugate complex 
$\bar z$, using the symbols $\partial$ and $\bpart$ to denote the 
partial derivatives with respect to these two variables.  In these 
variables we have 
\be\label{eq:vortexpot} W={\hbar\over 2mi}\ln{z\over \bar z} \ee
for the vortex potential.  In these variables there is one 
dimensionless parameter in the equations of motion, which is
\be\label{eq:alpha}
\alpha={3\hbar\rho_n\over 4m\eta}\;;
 \ee
this is of order unity at high temperatures, but falls off at lower 
temperatures in $^4$He in a similar manner to $\rho_n/\rho$, since
$\hbar\rho/m\eta$ remains of order unity. We look for solutions of the
form
\be\label{eq:trans1} \rho_s w=\Re zf(z\bar z)\;,\ \ \rho_n\chi=\Im
zg(z\bar z)\; \ee 
since these have the symmetry which corresponds to uniform
flow.  We use $s$ to denote the variable $z\bar z=r^2/L^2$, and then the
components of velocity are given by
\[L\rho_s(v_x-iv_y)=f+sf'+s\bar f'e^{-2i\theta} \;,\]
\[
L\rho_n(u_x-iu_y)=g-f+s(g'-f') \]\be\label{eq:velocities} - s(\bar
g'+\bar f') e^{-2i\theta} \;.\ee

In these new variables eq.\ (\ref{eq:l2curl}) gives
\be\label{eq:normal2} 3s\bigl( s\frac{d^2}{ds^2} 
+2\frac{d}{ds}\bigr)^2 g=- i\alpha 
(2s^2f'''+7sf''+2f' ) \;, \ee 
and eq.\ (\ref{eq:l2div2}) gives 
\[ 4s\bigl( s\frac{d^2}{ds^2} +2\frac{d}{ds} 
-\frac{1}{4}\bigr) \bigl( s\frac{d^2}{ds^2} +2\frac{d}{ds}\bigr)f \]
\be\label{eq:super2} =-2i\alpha(sf''+f') +i\alpha(2s^2 g'''+5sg'') 
\;.\ee

It is helpful to consider the case $\alpha=0$, where the two equations
(\ref{eq:normal2}) and (\ref{eq:super2}) decouple.  In this case eq.\
(\ref{eq:normal2}) has four solutions, namely
\be\label{eq:vector0} g\propto s\;,\ \  g\propto \ln s\;,\ \ g\ {\rm 
constant}\;,\ \ g\propto 1/s\;.
\ee
The first of these gives a solution of the Navier--Stokes equation
with flow which is quadratic at large distances, the second is the
well-known solution that is logarithmic at large distances and gives
rise to a constant viscous force on any surface surrounding the axis,
the third gives uniform flow, and the fourth dipolar flow falling off
like $1/r^2$ at large distances. Equation (\ref{eq:super2}), in the 
limit $\alpha\to 0$, has solutions  
\[ f\ {\rm constant}\;,\ \  f\propto 1/s\;, \]
\be\label{eq:scalar0} f\propto s^{-1/2} 
I_1(\sqrt s)\;,\ \  f\propto s^{-1/2} K_1(\sqrt s) \;, \ee 
where $I_1,K_1$ are modified Bessel functions. The first two of
these, combined with the corresponding solutions of eq.\
(\ref{eq:vector0}), give uniform and dipolar superfluid flow, while
the third and fourth give counterflow of the normal and superfluid
components, growing and decreasing exponentially with distance from
the axis.  

Solutions of the actual equations (\ref{eq:normal2}) and
(\ref{eq:super2}) have the same character at large distances, but
some of them are modified at small distances from the axis.  The four  
solutions that are unchanged are $f'=0$, $g'=0$, $g'$ constant, and 
the solution $f=1/s,g=-1/s$.  Combinations of the first two give 
uniform superfluid or normal fluid flow, the third gives quadratically 
growing normal fluid velocity, and is not of interest to us, while the 
fourth gives dipolar superfluid flow.  The real part of the solution
\be\label{eq:univ} f=\rho_sv_0\;,\ \ g=\rho_s v_0\;, \ee 
gives uniform superfluid flow, no normal fluid flow.  The
temperature is a constant $T_0$ everywhere, and the other
thermodynamic variables are given by 
\be\label{eq:univ2} \delta p=
\rho_s v_0{\hbar\over mr}\sin\theta\;,\ \ \delta\mu=v_0{\hbar\over
mr}\sin\theta\;. \ee
This is an exact solution of the equations as we have approximated 
them, but would be perturbed by terms from the inhomogeneity in the 
equilibrium  value of $\rho_s$.  These perturbations lead to 
corrections of $w$ which are of order $1/r$, and to corrections of the 
pressure which are of order $1/r^3$.  The net force per unit length 
on a large cylindrical volume of fluid of radius $R$ due to pressure is
\be\label{eq:bernoulli}
-\int_0^{2\pi} \delta p\,\br\,d\theta=-{\pi\hbar\over m}\rho_s\hat 
z\times
\bv_0
\;,\ee
while the momentum flowing into the cylinder per unit length is
\be\label{eq:momdens}
-\int {\hbar\over m}\rho_s\hat \theta \,\bv_0\cdot \hat r\, d\theta
=-{\pi\hbar\over m}\rho_s\hat z\times \bv_0
\;.\ee
This gives the usual expression for the Magnus force on the vortex 
pinning center as the sum of these two contributions, equal to
\be\label{eq:magnus}
{\bf F}_M= {h\over m}\rho_s\hat z\times \bv_0
\;.\ee
This derivation is exactly the same as the derivation for the Magnus 
force in a classical compressible fluid.

Uniform normal fluid flow is given by
\be\label{eq:unin} f=0\;,\ \ g=\rho_n u_0\;. \ee
This gives uniform pressure, no momentum flow, and no viscous forces.  
The dipolar superfluid flow is obtained by taking the derivative of 
the vortex solution with respect to the position of the vortex core, 
which is, in our notation,
\be\label{eq:dipoles} f(s)\propto{1\over s}\;,\ \ g(s)=-f(s) \;. \ee

Various methods are available to find the remaining four solutions of 
the equation, or, rather, the three solutions that do not grow 
exponentially at large distances.  We have solved the equations 
numerically, we have studied the power series solutions in positive or 
negative powers of $s$, and we have looked at the systematic expansion 
of the solutions in increasing powers of $\alpha$ using a standard 
Green function technique for solving the inhomogeneous ordinary 
differential equations that arise.  Since $\alpha$ is indeed a small 
parameter for much of the temperature range of superfluid $^4$He, and 
the expansion in powers of $\alpha$ is more transparent than the other 
methods, we will outline this method, and mention some of the results 
from the other methods.

We start with a zero order approximation that is a linear 
superposition of the solutions in the limit $\alpha=0$, excluding 
those that increase faster than logarithmically at large distances, and
excluding the uniform superfluid flow, which we know about already, so 
that we have
\[g_0(s)=c_1+ c_2\frac{a^2}{s}+ c_3\ln (s/a^2) \;,\]
\be\label{eq:sol0}
f_0(s)= d_2\frac{a^2}{s}+ d_3\frac{1}{\sqrt s}K_1(\sqrt s) 
\;.\ee
Substitution of this in the right hand sides of eqs.\ 
(\ref{eq:normal2}), (\ref{eq:super2}) gives two functions $i\alpha 
G_0,i\alpha F_0$ which have the form
\[ G_0= -d_3\frac{1}{4}K_1'(\sqrt s)\;,\]
\be\label{eq:inhom0} F_0=-2(c_2+d_2)\frac{a^2}{s^2}
-c_3\frac{1}{s}  +d_3\frac{1}{2\sqrt{s}}
K_2'(\sqrt s)  \;, \ee  
where $aL=r_0$ is the distance at which short distance boundary conditions will
be imposed. This leads to the inhomogeneous equations
\[ 3s\bigl( s\frac{d^2}{ds^2} +2\frac{d}{ds}\bigr)^2 g_1= i\alpha G_0 
\;, \] 
\be\label{eq:iterate} 4s\bigl( s\frac{d^2}{ds^2} +2\frac{d}{ds} 
-\frac{1}{4}\bigr) \bigl( s\frac{d^2}{ds^2} +2\frac{d}{ds}\bigr)f_1 
=i\alpha F_0
\;.\ee
Particular integrals of these inhomogeneous equations can be obtained by a 
standard Green function technique, in the form
\[
g_1(s)=i\alpha\int_{a^2}^\infty K_g(s,t)G_0(t)\,dt \;,\]
\be\label{eq:green1}
f_1(s)=i\alpha\int_{a^2}^\infty K_f(s,t)F_0(t)\,dt \;,
\ee 
and solutions $c_1'+c_2'a^2/ s, d_2'a^2/ s +d_3' K_1(\sqrt
s)/\sqrt s$, should be added to the particular integral so that the
boundary conditions at $s=a^2$ are satisfied.  We keep the coefficient
of $\ln(s/a^2)$, the dominant term at large distances, constant
during the iterative procedure. The Green functions
$K_f,K_g$ can be written as
\[ K_f(s,t)=\cases{1/s-(2/ \sqrt{st})K_1(\sqrt s)I_1(\sqrt 
t)&if $t<s$,\cr
1/t-(2/ \sqrt{st})I_1(\sqrt s)K_1(\sqrt t)&if $t>s$,}\]
\be\label{eq:green2}
K_g(s,t)=\cases{(2\ln t-t/s)/6 &if $t<s$,\cr
(2\ln s-s/ t)/6&if $t>s$,\cr} \ee 
but it is actually convenient sometimes to use
other forms which differ from these by a solution of the homogeneous
equation.

The process can be repeated by substituting $f_1,g_1$, etc., in the right 
hand sides of eqs. (\ref{eq:normal2}), (\ref{eq:super2}) to generate 
$F_1,G_1$, and so to find the second order corrections in $f,g$.

All the solutions generated in this way have series solutions in 
ascending powers of $s$, with the leading power $n_0$ given by the
solutions of the complex indicial equation
\be\label{eq:index} 12 n_0^2(n_0^2-1) +6i\alpha n_0^2
-\alpha^2 (4n_0^2-1) =0 \;.\ee
The solution generated by a combination of $c_2,d_2$ has a power 
series solution in descending powers of $s$, with $1/s$ as the leading 
term. The solution generated by $c_3$ is $\ln s$ times a 
descending power series, and is the solution which leads to a net 
viscous force at large distances.  The solution generated by $d_3$ 
has a singular point at infinity.

 \section{Boundary conditions and solutions}\label{sec:bc}
\setcounter{equation}{0}

In the previous section we have shown how linearized solutions of the 
two-fluid equations in the presence of a straight superfluid vortex can 
be found.  We must combine these solutions in a way that satisfies 
appropriate boundary conditions.  We already know that superfluid flow 
gives rise to a Magnus force whose magnitude depends on the superfluid 
velocity at large distances.  It remains to find the corresponding 
solution for the normal fluid component.  As with a classical fluid, we 
cannot satisfy boundary conditions simply with a flow which is 
asymptotically uniform, but we need the logarithmically growing term. 
Four relations between the other five coefficients  must be determined
by boundary conditions on some inner boundary.

If the fluid is rotating around a wire of circular cross-section, 
radius $r_0=aL$, as in the Vinen experiment \cite{vinen61}, it is 
natural to take both components of normal fluid velocity, and the 
radial component of superfluid velocity, to be zero at $r_0$, or, in 
the notation of eq.\ (\ref{eq:trans1}),
\[
2a^2f'(a^2) +f(a^2)=0\;,\ \  g(a^2)=0\;,\]
\be\label{eq:sticky} 2a^2g'(a^2)-f(a^2)=0\;.\ee
 An additional condition must be imposed, such as that
the temperature is  uniform on the boundary (alternatively, it might
have been more realistic to take the heat flux to be zero at the
boundary).  From eq.\ (\ref{eq:l2temp}) the condition for constant
temperature at $r=aL$ is
\[
\frac{\rho_s\hbar}{maL}\bigl(\nabla^2w +\frac 
{1}{a^2L^2}\partial_\theta^2w\bigr) 
+\frac{4\rho_s\eta}{3\rho_naL}\nabla^2\partial_\theta w \]
\be\label{eq:temp2} +\bigl(\eta\nabla^2+ 
\frac{\rho_n\hbar}{ma^2L^2}\partial_\theta\bigr)\partial_r\chi =0 \;,
\ee or, in terms of $f$ and $g$, and making use of eq.\
(\ref{eq:sticky}),
\be\label{eq:temp3}
3(2a^4g'''+7a^2g''+2g')  =4(1-i\alpha)(a^2f''+2f') \;.\ee
Only the terms for which $\nabla^2\chi$ and $\nabla^2 w$ are nonzero
contribute to this.

Our main concern in this work is to what extent the normal fluid flow 
past a vortex can generate a transverse force.  The direction of the 
viscous force is determined by the phase of the coefficient of the  term 
in $g$ which is logarithmic in $s$ for large $s$, while the direction of 
flow is determined also by the constant term.  The existence
of a transverse force depends on the complexity of the  solutions of
eqs.\ (\ref{eq:normal2}) and (\ref{eq:super2}), imposed  either by the
form of the equations or by the boundary conditions.  Numerical solution
shows that in this case (no tangential normal fluid flow or temperature
gradient at the boundary) there is a transverse force due to normal
fluid flow, and that this transverse force is proportional to
$\rho_n$ and more or less independent of $\eta$. This transverse force is
small when the radius of the inner boundary is significantly greater
than $L$, but rises as $a$ approaches unity. 

In Appendix \ref{sec:firstorder} some of the details of the solution
of the equations with these boundary conditions to first order in
$\alpha$ and lowest nonvanishing order in $1/a$ are shown.  To zero
order in $\alpha$ and for $a>>1$, eq.\ (\ref{eq:bc0}) gives
\be\label{eq:bcsoln} c_1= \bigl[-1+ \frac{3}{2a}\bigr]c_3 \;,\ee
which shows that the solution is similar to the solution of the
classical problem, and the coupling between normal and superfluid
components simply reduces the effective radius of the cylinder by an
amount close to $3L/4$. 

There are imaginary corrections to this solution to first order in
$\alpha$, coming both from the combined effect of the Green functions in
eq.\ (\ref{eq:green1}), and from the boundary conditions.  The
detailed work in Appendix \ref{sec:firstorder} shows that the most
important terms for $a>>1$ are of order $i\alpha/a^2$.  When the
coefficient $c_3$ of the logarithmic term in $\bu$ is kept equal to
a constant $C_3$ in the iterative procedure, the coefficient of the
constant term is, from eq.\ (\ref{eq:c1}),
\be\label{eq:c12} C_1\approx c_1+c_1'=\bigl[- 1+
\frac{3K_0(a)}{2aK_1(a)}\bigr]C_3 - \frac{2i\alpha }{a^2}C_3  \;.
\ee

In general the normal fluid velocity at large distances should have the
form
\be\label{eq:ularge} L\rho_n (u_x-iu_y)= C_3[\ln
(s/a^2)+1]-\bar C_3e^{-2i\theta}+C_1 \;.
\ee
Because of the logarithmic term, this does not tend to uniform flow
at large distances, which is the basis of the well-known Stokes
paradox for slow viscous flow past a cylinder.  The resolution of the
Stokes paradox was provided by Oseen in 1910 (see refs.\
\cite{batchelor67} and \cite{lamb45}), who pointed out that at
sufficiently large distances the nonlinear inertial term,
$\rho_n\bu\cdot\nabla\bu$, in the equation of motion, overwhelms the
viscous term, $\eta\nabla^2\bu$, however small the fluid velocity
$\bu$ is. This was discussed in the context of the drag on
superfluid vortices by  Hall and Vinen \cite{hallvinen56} and by
Vinen \cite{vinen57}, and Appendix \ref{sec:hallvinen} shows how our
discussion relates to this early work. In the region where these
nonlinear terms are important there is negligible coupling between
the superfluid and normal components, so the classical theory can be
taken over with little modification.

Oseen proposed replacing the nonlinear term $\rho\bu\cdot \nabla\bu$ in
the Navier--Stokes equation by $\rho{\bf U}\cdot \nabla\bu$, where $\bf U$
is a constant vector equal to the asymptotic velocity of the fluid.  This
gives the leading term in a small Reynolds number asymptotic
expansion, uniform in $r$, and is a good approximation in the regions
where the nonlinear term is more important than the viscous term. The
vorticity for the normal fluid then satisfies the  equation 
\be\label{eq:oseen}
\rho_n{\bf U}\cdot {\bf curl}\, \bu =\eta \nabla^2 {\bf
curl}\,\bu\;,\ee
which follows from the curl of eq.\ (\ref{eq:forces}) by dropping all
terms involving the divergence of $\bu$.  The general solution of this
equation with the right asymptotic behavior, with $\bf U$ in the $x$
direction, is 
\be\label{eq:oseensol1}
{\bf curl}\,\bu =e^{kx}\sum_n p_nK_n(kr)e^{in\theta} \;, \ee
where the $p_n$ are arbitrary coefficents, $K_n$ is a modified Bessel
function, and 
\be\label{eq:defk}
k= \frac{\rho_nU}{2\eta}\;. \ee 

Lamb \cite{lamb45} gave an explicit solution for $\bu$, and a simple
generalization of this solution to allow for a transverse component of
the force is
\[ u_x-iu_y=U -A e^{kx}K_0(kr)-\bar Ae^{-i\theta}[e^{kx}K_1(kr)-1/kr] \]
\[ = U-A e^{kL(z+\bar z)/2}
K_0(kL\sqrt {z\bar z}) \]
\be\label{eq:oseensol2} -\bar A\sqrt{\bar z\over z} e^{kL(z+\bar z)/2}
K_1(kL\sqrt {z\bar z}) +\frac{\bar A}{kLz}  \;.\ee
It can be verified that this indeed gives incompressible flow whose curl
satisfies the Oseen equation, since we have
\[
\frac{2i}{L}\frac{\partial}{\partial\bar z}(u_x-iu_y)={\bf
curl}\,\bu+i\,{\bf div}\,\bu \]
\be\label{eq:oseensol3} = 2ke^{kx} [\Im A K_0 -\Im
(Ae^{i\theta}) K_1]\;.\ee
The vorticity given by this expression vanishes for $kr>>1$, except in a
parabolic wake bounded by $\theta^2\approx 2/kr$; outside the wake the
flow is a combination of dipolar and vortex motion.  A general
discussion of this method has been given by Proudman and Pearson
\cite{proudman57}.

The leading term in the expansion of eq.\ (\ref{eq:oseensol2}) in
powers of $k$ is
\be\label{eq:smallk}
u_x-iu_y \approx U +A \ln(kL|z|/2)+A\gamma -\frac{\bar A}{2}
-\frac{\bar A\bar z}{2z}\;, \ee 
where $\gamma$ is Euler's constant. In the range $L<<r<<1/k$ this should
match the form given in eq.\ (\ref{eq:ularge}), and this requires
\be\label{eq:fit1} A=2C_3/L\rho_n\;,\ee
and
\be\label{eq:fit2}
L\rho_nU =  C_3[2\ln(2/kLa)-2\gamma ] +C_3+\bar C_3 + C_1 
 \;. \ee
The complex ratio $C_1/C_3$ is determined by the solution of the
differential equations (\ref{eq:normal2}) and (\ref{eq:super2}), and
by the boundary conditions, so the phases of $C_1,C_3$ are determined
by the imaginary part of eq.\ (\ref{eq:fit2}), and the amplitudes are
related to the asymptotic normal fluid velocity $U$ by the real
part.  In the case that we have studied in detail this is given in
eq.\ (\ref{eq:c12}).

 Substitution of eq.\ (\ref{eq:ularge}) in eq.\ (\ref{eq:forces}) gives
the change of pressure and the viscous stress tensor in the region
$L<<r<<1/k$ as
\[\delta p=-{4\eta\over rL\rho_n} (\Re C_3\cos\theta-\Im
C_3\sin\theta)\;,\]
\be\label{eq:viscforce}
S_\eta={4\eta\over rL\rho_n}(\hat r\hat r-\hat\theta\hat\theta)
(\Re C_3\cos\theta-\Im C_3\sin\theta) \;.\ee 
Pressure and viscous stress each contribute half the net force, so that
the total force per unit length due to normal fluid flow is 
\be\label{eq:fu} \bF_n= {8\pi\eta\over L\rho_n} (\Re C_3\hat x-\Im
C_3\hat y) \;, \ee 
in agreement with the results for an ordinary fluid quoted in
hydrodynamics text books \cite{batchelor67}.  Equations
(\ref{eq:oseensol2}) and (\ref{eq:fit1}) show that the normal fluid has
circulation
\be\label{eq:normalcirc}
\kappa_n= \frac{4\pi\Im C_3}{kL\rho_n} \ee
in the region $r>>1/k$. Comparison of this with eqs.\ (\ref{eq:fu}) and
(\ref{eq:defk}) shows that the transverse component of the force is
given by
\be\label{eq:tforce1}
F_t= -\rho_n\kappa_n U\;.\ee
This is in full agreement with the equations derived in TAN
\cite{tan96}, but not with the discussion, since it was assumed that
the normal fluid circulation is zero.  In fact, as we shall show,
$\kappa_n$ goes to zero as $U$ goes to zero, but only as fast as
$1/|\ln U|^2$.

The real parts of eqs.\ (\ref{eq:fit2}) and
(\ref{eq:fu}) combined with the approximation (\ref{eq:bcsoln}) gives 
\be\label{eq:fnl} {{\bf F}_n\cdot \hat x\over U}\approx {8\pi\eta \over
2\ln (4\eta/\rho_nUa) -2\gamma +1+3/2a} \;,\ee
which is the standard result from classical hydrodynamics, apart from a
small correction of order $1/a$ in the denominator.  Combination of the
imaginary parts of eqs.\ (\ref{eq:c12}) and (\ref{eq:fit2}) gives
\be\label{eq:imre} \Im C_3\approx
\frac{4\alpha/a^2}{2\ln(4\eta/\rho_nUa)-1+3/2a}\Re C_3 \;,
\ee
so that, after substitution of the value of $\alpha$ from eq.\
(\ref{eq:alpha}), the transverse force is given by
\be\label{eq:fnt} {\bf F}_n\cdot \hat y \approx 
-\frac{3\rho_n(h/m)}{a^2[\ln
(4\eta/\rho_nUa) -\gamma/2 ]^2}U
\;.\ee 
which agrees with the result written down in eq.\ (\ref{eq:tnforce}). 
As the cut-off $R_c=4\eta/\rho_nUa$ goes to infinity both these
components go to zero, the transverse force faster than the longitudinal
force.   For  any finite value of the cut-off these corrections of order
$1/\ln R_c$ and $1/(\ln R_c)^2$ will be needed.  Our numerical results
agree with these expressions (\ref{eq:fnl}) and (\ref{eq:fnt}).

This calculation shows that even a solid core can give rise to a
transverse force which is proportional to the normal fluid density,
and its ratio to the normal fluid velocity close to the core can be
quite large if the core radius is comparable with $L$ ($a$ of order
unity), however, its ratio to the asymptotic value of the normal
fluid velocity is reduced by a factor proportional to $1/(\ln
R_c)^2$.

 As was observed by Hall and Vinen, the actual cut-off may be reduced to
lower values, if a nonzero angular velocity produces a finite spacing of
vortices, producing a situation analogous to the array of cylindrical
obstacles considered for a viscous fluid by Sewell \cite{sewell10}, or
if there is a nonzero  frequency $\nu$ of oscillation of the normal
fluid relative to the vortex, in  which case a cutoff of magnitude
\be\label{eq:cutoff2} R_c\approx \sqrt{\eta\over \rho_n\nu} \;,\ee
is appropriate.

\section{Free vortices and the Iordanskii force}\label{sec:iordanskii}
\setcounter{equation}{0} For a free vortex core, or one bound by a core
whose size is much less  than the mean free path of excitations, then we
should take account of  the effect of the superfluid flow on
noninteracting excitations, in  accordance with the discussions of
Lifshitz and Pitaevskii \cite{pitaevskii58},  Iordanskii
\cite{iordanskii64}, and
others \cite{sonin76,nielshede95}.  According to these works the flow of
phonons or rotons past a stationary  vortex produces a transverse force
equal to
\be\label{eq:transforce}
{\bf F}_t=-{\rho_n h\over m}\hat z\times \bu_c\;,\ee
where $\bu_c$ is the normal fluid velocity at a distance of the 
order of a mean free path from the vortex core.  In the phonon dominated 
regime there is a longitudinal force
\be\label{eq:longforcep}
{\bf F}_{lp}= {\rho_n h\over m} {k_BT\over 
mc^2} {15\pi \zeta(5)\over 4\zeta(4)}\bu_c\;,\ee
where $c$ is sound velocity and $\zeta$ denotes the Riemann
$\zeta$-function. In the roton dominated region the  longitudinal force is
\be\label{eq:longforcer}
{\bf F}_{lr}\approx {\rho_n h\over m} 3\pi\sqrt{{2\pi\mu\over
m}{p_0^2\over mk_BT}} \ln\bigl(\sqrt{2\mu k_BT}\frac{2L}{h} \bigr)
\bu_c\;,\ee
where $\mu$ is roton mass and $p_0$ is the magnitude of roton momentum.  
For helium in the phonon-dominated region, the longitudinal force due to
normal fluid flow is about 40\% of the transverse force, while in the
roton dominated region the longitudinal force is considerably larger than
the transverse force.

A detailed calculation of what happens in this case is not easy to
undertake, since we have no satisfactory theory of the transition region
between the collisionless region  of the order of a mean free path in
radius and the hydrodynamic region beyond a few mean free paths from the
vortex core.  The best we have been able to do is to join one region
abruptly to the other. Qualitatively the same thing will happen that happens for a solid core, which is that the growth of the logarithmic term in the normal fluid velocity will lead to an increasing alignment of the direction of 
normal fluid flow with the direction of the force at larger distances 
from the core, until the Oseen or other cut-off radius $R_c$ is 
reached.   Indeed, with a solid core whose radius $r_0$ is comparable 
with the mean free path $L$, two-fluid hydrodynamics, although it is 
not strictly applicable, predicts a transverse force comparable with 
the Iordanskii force.The phonon-dominated regime is a little different, because the transverse force is initially somewhat larger than the longitudinal force,
so, even after the logarithmic growth of the normal fluid comes into
effect, there is still a significant angle between the direction of the
force and the direction of flow.

\section{Discussion}\label{sec:disc} \setcounter{equation}{0}

In this paper we have investigated how to combine the TAN expression
relating the transverse force on a vortex to the circulation of momentum
round the vortex to the usual results based on a combination of
two-fluid hydrodynamics with the Lifshitz--Pitaevskii--Iordanskii
expressions for
the force on a free vortex in the presence of superfluid and normal
flow.  We have found that for a single stationary vortex in an infinite
superfluid the TAN relation between transverse force and circulation of
momentum is correct --- this is not surprising, since the relation
depends on little more than conservation of momentum and the
independence of normal and superfluid flows at large distances from the
vortex.  The assumption that is wrong in ref.\ \ref{tan96} is that there
is no normal fluid circulation, which is only literally true when the
normal fluid is at rest.  Application of the methods of Oseen and Lamb
to the two-fluid situation shows that normal fluid flow with speed $U$
round a vortex  induces a circulation proportional to  $\rho_n h U/ 
m[\ln(4\pi\eta U/\rho_na)]^2$.

    For an isolated vortex this factor proportional to the inverse
square of a logarithm will be rather small, and so the transverse force
due to normal fluid flow will be small compared with the transverse
force due to superfluid flow.  There is a possible modification of this
conclusion when the normal fluid component is primarily phonons, but then
the normal fluid density is very small relative to the superfluid
density, so the transverse force due to the normal fluid flow is still
going to be small.

  Similar conclusions have been reached in the recent paper by Sonin
\cite{sonin00}.

\section*{Acknowledgements}

This work was supported in part by the US NSF under grant number
DMR-9813932, by a Sarah Moss Fellowship, and by the Research
Corporation.  We have had many useful discussions with Ping Ao,
Moo-Young Choi, Qian Niu, Edouard Sonin, Boris Spivak, Michael Stone,
Jian-Ming Tang and Carlos Wexler. Jian-Ming Tang has been particularly 
helpful in looking for errors at a late stage in drafting this paper. 
We are grateful for the hospitality of the Aspen Center for Physics and 
of the Isaac Newton Institute of Cambridge University.

\begin{appendix}
\def\theequation{\Alph{section}\arabic{equation}}\setcounter{equation}{0}
\section{First order corrections}\label{sec:firstorder}

We work out the first order corrections to the flow, taking the specific
case of the boundary conditions in which there is no normal fluid flow and
constant temperature at $r=aL$, where $a$ is large, as given in eqs.\
(\ref{eq:sticky}) and (\ref{eq:temp3}).  

To zero order the functions $g$ and $f$ are given by\onecolumn
\be g=c_1+c_2\frac{a^2}{s}+c_3\ln\frac{s}{a^2}\;,\ \ 
 f=d_2\frac{a^2}{s}+d_3\frac{1}{\sqrt{s}}K_1(\sqrt{s})\;.\ee
The four boundary conditions give
\be d_2=d_3K_1'(a)\;,\ \ c_2+c_1=0\;,\ \ 
 -2c_2+2c_3=d_2+d_3\frac{1}{a}K_1(a)\;,\ \ 
 3c_3=-(1-i\alpha)d_3 aK_1(a)\;. \ee
These equations have solutions
\be\label{eq:bc0} c_1= \bigl[-1+
\frac{3K_0(a)}{2aK_1(a)(1-i\alpha)}\bigr]c_3=-c_2
\;,\ \ 
 d_2= -\frac{3K_1'(a)}{aK_1(a)(1-i\alpha)}c_3\;,\ \ 
d_3= -\frac{3}{aK_1(a)(1-i\alpha)}c_3 \;.
\ee
 
 Substitution of eq.\ (\ref{eq:inhom0}) in eqs.\ (\ref{eq:green1}) and
(\ref{eq:green2}) gives 
\be\label{eq:iter1g}
 g_1(s)= d_3 \frac{i\alpha}{24}\int_{s}^\infty \bigl[
2\ln\frac{t}{s} -\frac{t^2-s^2}{st}\bigr] K_1'(\sqrt t)\,dt
=d_3\frac{i\alpha}{3}\int_s^\infty\frac{t+s}{t^{5/2}}
K_1(\sqrt{t})\,dt  \;,\ee where three integrations by parts have been
performed, and
\[ f_1(s)=i\alpha(c_2+d_2)\Bigl[\int_s^\infty
\frac{2a^2(t-s)}{st^3}\,dt +\int_{a^2}^s \frac{4a^2}{s^{1/2}t^{5/2}}
K_1(\sqrt s)I_1(\sqrt t)\,dt +\int_s^\infty \frac{4a^2}
{s^{1/2}t^{5/2}} I_1(\sqrt s)K_1(\sqrt t)\,dt\]\[ +\frac{8}{s^{1/2}a^2}
K_1(\sqrt s)I_1(a)\Bigr]  +i\alpha c_3\Bigl[-\int_{a^2}^s
\frac{1}{st}\,dt +\int_{a^2}^s \frac{2}{s^{1/2}t^{3/2}} K_1(\sqrt
s)I_1(\sqrt t) \,dt +\int_s^\infty \frac{2}{s^{1/2}t^{3/2}} I_1(\sqrt
s)K_1(\sqrt t)\,dt\]
\be\label{eq:iter1f}+\frac{4}{s^{1/2}a^2} K_1(\sqrt s)I_1(a)\Bigr]
+i\alpha d_3\Bigl\{ -\int_s^\infty \frac{t-s}{2st^{3/2}} K_2'(\sqrt t) \,dt 
 +\int_s^\infty \frac{1}{s^{1/2}t}
\bigl[K_1(\sqrt s)I_1(\sqrt t) -I_1(\sqrt s)K_1(\sqrt t)
\bigr]  K_2'(\sqrt t)  \,dt\Bigr\} \;. 
\ee 
To leading orders in $a,s$, these give
\be\label{eq:itera} g_1(s) \approx d_3i\alpha \sqrt{\pi\over 2}\Bigl(
\frac{4}{3s^{5/4}}-
\frac{25}{6s^{7/4}}\Bigr)e^{-\sqrt s} \;,\ \ 
 f_1(s) \approx i\alpha\Bigl[(c_2+d_2)
\frac{a^2}{s^2} -c_3\frac{1}{s}\ln(s/a^2) -d_3\sqrt{\pi\over
2}\frac{1}{s^{5/4}}e^{-\sqrt s}\Bigr]\;.
\ee

Now we substitute the expressions
\be g_1(s)+c_1'+c_2'\frac{a^2}{s}\;,\ \ f_1(s)+d_2'\frac{a^2}{s}+d_3'
\frac{1}{\sqrt{s}}K_1(\sqrt s)\;,\ee
in the boundary conditions (\ref{eq:sticky}) and (\ref{eq:temp3}) to get
\[
c_1'=\frac{K_0(a)}{aK_1(a)} \bigl[-\frac{3}{2}(2a^6g_1'''+7a^4g_1''
+2a^2g_1')+2(a^4f_1''+2a^2f_1')\bigr] +a^2(f_1'-g_1')+f_1-g_1 \;,\]
\be\label{eq:bc1} c_2'=-c_1'-g_1\;,\ \
d_3'K_0(a)=2c_2'+2a^2(f_1'-g_1')+2f_1\;,\ \ 
 d_2'=d_3'K_1'(a)+2a^2f_1'+f_1\;.\ee 
The coefficient
$c_1'$ is the most important one, since, with
$c_1,c_3$ it dominates the behavior for large $s$.  In the
expression for $c_1'$ it can be seen that the only terms giving
contributions of order $i\alpha/a^2$ are the terms proportional to
$a^3g_1'', a^2g_1',a^3f_1'', a^2f_1', f_1$, while the term proportional to
$a^5g_1'''$ gives a contribution of order $i\alpha/a$ which
cancels with a similar term in $c_1$.  There is no contribution
to this order from the terms in $f_1$ proportional to $d_2$ or
$d_3$.  With the substitution of eqs.\ (\ref{eq:bc0}) and
(\ref{eq:itera}) in this we get
\be\label{eq:bc12}
\frac{c_1'}{i\alpha}\approx d_3 \frac{K_0(a)}{2}-(c_2+c_3)
\frac{1}{a^2} +d_3 \sqrt{\pi\over 2}
\frac{2}{3a^{3/2}}e^{-a}  \approx -c_3 \bigl[
\frac{3K_0(a)}{2aK_1(a)}+\frac{4}{a^2}\bigr] \;.
\ee
Combined with eq.\ (\ref{eq:bc0}) this gives 
\be\label{eq:c1}
\frac{c_1+c_1'}{c_3}\approx \bigl[- 1+
\frac{3K_0(a)}{2aK_1(a)}\bigr] -4 
\frac{i\alpha}{a^2}  \;.
\ee
 The last term in
eq.\ (\ref{eq:c1}) gives the imaginary part of the coefficient,
which leads to a velocity at large distances which is not in the
same direction as the force.  
\twocolumn

\section{Relation to the Hall--Vinen theory}\label{sec:hallvinen}
\setcounter{equation}{0}

As is clear from earlier parts of this paper, some of the problems that
we have addressed were foreseen by Hall and Vinen \cite{hallvinen56}
(HV) and by Vinen \cite{vinen57},  and it may help the reader if we
examine the extent to which the results derived in this earlier work
accord with those derived here.   The earlier work referred to a vortex
with a core of microscopic size (much smaller than the excitation mean
free path);  furthermore,  as we have noted,  it did not include any
transverse force due to flow of the normal fluid,  such a force having
been ruled out,  incorrectly,  by a supposed symmetry of the scattering
of excitations by the vortex.  However,  it is easy to modify the
earlier work by the addition of a transverse force,  as we now show.

 In the presence of a transverse component of the force,  equation (2) of
HV must be generalized to
\begin{equation}f_{\parallel}=Dv_{R\parallel}-D'v_{R\perp}\;,\
\ f_{\perp}=Dv_{R\perp}+D'v_{R\parallel};\end{equation}
 The symbols ${\parallel}$  and ${\perp}$  refer to directions parallel
and perpendicular to the asymptotic normal fluid velocity,  $U$ (i.e.
$f_{\parallel}={\bf F}_n\cdot\hat{x}$ and
$f_{\perp}={\bf F}_n\cdot\hat{y}$ in the notation of section 4 ),  and
${\bf v}_{R}$ is the velocity of the normal fluid close to the
(stationary) vortex.  Equation (4) of HV still holds and gives the
relationship between $U$ and ${\bf v}_{R}$
\be\label{eq:hv3}U-v_{R\parallel}=\frac{f_{\parallel}}{E}\;,\ \ 
-v_{R\perp}=
\frac{f_{\perp}}{E}\;,\end{equation}
where
$E$ is given by equation (5) of HV for the case of finite frequencies. 
The parameter $L$ in $E$ is roughly the mean free path of the excitations
constituting the normal fluid,  as in earlier parts of this paper.  In
the case of zero frequency,  with which this paper is concerned,  the
parameter $\lambda$ in equation (5) of HV must be replaced by
\cite{vinen57} $\rho_{n}U/2\eta$,  so that the logarithmic term in
equation (5) of HV becomes $\ln{R_e}$, where $R_{e}$ is the Reynolds
number $L\rho_{n}U/4\eta$. When $U$ is sufficiently small the magnitude
of this (negative) logarithmic term is much larger than unity, and then
$E$ can be written in the approximate form
\begin{equation}\label{eq:hv4}
E\approx\frac{4\pi\eta}{\ln R_e}\;.\end{equation}

 From these equations we can derive the relations between the components
of the force on the vortex and $U$,  analogous to equations
(\ref{eq:fnl}) and (\ref{eq:fnt}) in section 4.
\[\frac{f_{\parallel}}{U}=E-\frac{(D+E)E^{2}}{(D+E)^{2}+D'^{2}}\;; \]
\begin{equation}\frac{f_{\perp}}{U}=\frac{E^{2}D'}{(D+E)^{2}+D'^{2}}
\;.\end{equation} 
For sufficiently small values of $U$ we must have $D\gg{E}$. It follows
that the leading term in $f_{\parallel}/U$ is proportional to
$[\ln{R_{e}}]^{-1}$ ,  while the leading term in  $f_{\perp}/U$ is
proportional to $[\ln{R_{e}}]^{-2}$. Equations (\ref{eq:fnl}) and
(\ref{eq:fnt}) in section 4 exhibit the same feature.  According to HV
the feature has its origin in the form of equations (\ref{eq:hv3}), 
which give the response of the normal fluid to the line force
${\bf f}$.  The cut-off  in the solution of the linearized Navier-Stokes
equation that is required at large  distances (section 1) forces this
response to decrease logarithmically as $U$ decreases and the
corresponding cut-off distance increases.
\end{appendix}

\end{document}